# Does tumor growth follow a "universal law" ?

Caterina Guiot[*,†], Piero Giorgio Degiorgis[†,‡], Pier Paolo Delsanto[†,§], Pietro Gabriele[¶] and Thomas S. Deisboeck[#,**]

*Dip. Neuroscience, Università di Torino, Italy and [†]INFM, sezioni di Torino Università e Politecnico, Italy ; [‡] S.I.A. SpA, Torino; [§] Dip. Fisica, Politecnico di Torino, Italy; [¶] IRCC Candiolo, Torino; [#] Complex Biosystems Modeling Laboratory, Harvard-MIT (HST) Athinoula A. Martinos Center for Biomedical Imaging, HST-Biomedical Engineering Center, Massachusetts Institute of Technology, Cambridge, MA 02139 and [**] Molecular Neuro-Oncology Laboratory, Harvard Medical School, Massachusetts General Hospital, Charlestown, MA 02129, USA.*

**corresponding author:**

Caterina Guiot
Dip. Neuroscienze
30, C. Raffaello
10125  Torino
tel: +39.11.670.7710
fax: +39.11.670.7708
e-mail: caterina.guiot@unito.it

**A general model for the ontogenetic growth of living organisms has been recently proposed. Here we investigate the extension of this model to the growth of solid malignant tumors. A variety of *in vitro* and *in vivo* data are analyzed and compared with the prediction of a 'universal' law, relating properly rescaled tumor masses and tumor growth times. The results support the notion that tumor growth follows such a universal law. Several important implications of this finding are discussed, including its relevance for tumor metastasis and recurrence, cell turnover rates, angiogenesis and invasion.**

A better understanding of the growth kinetics of malignant tumors is of paramount importance for the development of more successful treatment strategies. Given the lack of clinical data at non-symptomatic stages, it has been conjectured, that in most solid malignant human tumors two or three decades elapse between the first carcinogenic stimulus and the clinical emergence of the neoplasm (Tubiana, 1986). Since a tumor is clinically detectable with conventional diagnostic tools at approximately 1 cm$^3$ in volume, representing a population of about 1 billion cells, some 30 cell doublings from the progenitor cancer cell must occur in order to reach this 'diagnostic' stage. Assuming typical volume doubling times of about 100 days, this scenario corresponds to a preclinical time of roughly 8 years . In the next 10 volume doublings up to about 10$^3$ cm$^3$ in size (which is reportedly lethal in primary breast cancer by Retsky, 1997), the clinical history of the tumor passes through a microscopic, avascular growth phase, followed by angiogenesis, necessary to sustain a macroscopic size (Folkman, 1971).Continuous tumor progression leads then eventually to local tissue invasion and metastasis, depending on the particular cancer type.

A recent paper from West et al (2001) shows that, regardless of the different masses and development times, mammals, birds, fish and molluscs all share a common growth pattern. Provided that masses and growth times for the different organisms are properly rescaled, the *same* universal exponential curve fits their ontogenetic growth data. The authors explain this phenomenon with basic cellular mechanisms (West et al, 2002), assuming a common *fractal* pattern in the vascularization of the investigated taxa.

Since rapid volumetric growth and neovascularization are also hallmarks of solid malignant tumors, it is intriguing to investigate whether tumor growth follows the same *universal* curve. For this purpose we rescale tumor sizes and growth times according to West's recipe (West et al, 2001), which we briefly recall here. Instead of the actual mass, *m*, they use the ratio $r = (m/M)^{0.25}$, where *M* is the asymptotic mass for the taxon; *r* relates to the relative proportion of total energy expenditure, which is required to ensure maintenance. Likewise, instead of the actual time, *t*, these authors use the rescaled dimensionless time

$$\boldsymbol{t} = \frac{1}{4} aM^{-0.25} t - \ln(1 - (\frac{m_0}{M})^{0.25}) = \boldsymbol{a} \, r_0 t - \ln(1 - r_0)$$
[1]

where $r_0 = (m_0/M)^{0.25}$, $m_0$ is the mass at birth, $\alpha = 0.25 \, a \, m_0^{-0.25}$ and $a$ is a parameter, which is roughly constant within a taxon and proportional to the organism's metabolic rate across taxa. The values of $a$, $m_0$ and $M$ are reported for many different species in West et al (2001). Using the variables $\boldsymbol{t}$ and $r$, this 'universal' growth law follows:

$r = 1 - exp(-\boldsymbol{t})$ [2]

In our case, $m_0$ and $M$ are the initial and final masses of the tumor, and $a$ is a parameter expected to be related to the tumor's characteristics (e.g., its ability to metastasize or invade, or its affinity for nutrients uptake). Since the definition of the parameters is non-trivial, a multistep fitting procedure is adopted for their determination. From equations [1] and [2] we obtain:

$y = -\boldsymbol{a} \, r_0 \, t + y_0$ [3]

where $y = \ln(1-r)$, $y_0 = \ln(1-r_0)$. Available data normally provide $m$ as a function of $t$ during the tumor's growth phase. For our fitting procedure, the lowest and largest values of $m$ are first assumed for $m_0$ and $M$, respectively, thus yielding a preliminary estimate for $y_0$ and $\boldsymbol{a}$. Then $m_0$ and $M$ are allowed to respectively decrease and increase, and $y_0$ and $\boldsymbol{a}$ reestimated until a best-fit (consistent with the available biological infomation) is obtained.

For our analysis, we have used available data from the literature, spanning *in vitro* experiments (multicellular tumor spheroids (Chignola et al, 2000, Nirmala et al, 2001) as well as *in vivo* data (both, from animal models (Steel, 1977; Cividalli et al, 2002) and patients(Norton, 1988; Yorke et al, 1993). The results are presented in Figures 1, 2 and 3, respectively, and plotted against equation [2]. The data fit the universal growth curve very well. Table 1 presents an estimate of the relevant parameters and, further supporting our claim, shows very high $R^2$ correlation coefficients between actual and fitted data.

In the following, we will briefly discuss possible implications of our conjecture that tumor growth also follows a universal law.

**a)** *tumor metastasis and recurrence.* Some forty years ago Romsdahl et al. (1961) had proposed that, upon reaching a certain "*critical*" volume of the primary tumor, metastatic dissemination would start. Tubiana (1986) evaluated this threshold for breast cancer, and showed that it critically depends on the axillary lymph-node involvement. The hypothesis that metastatic behaviour could be associated with a universal growing phase of the tumor is corroborated by experimental evidence. For example, data from C3Hf/Sed mice inoculated with fibrosarcoma (FSall) and squamous cell carcinoma (SCCVII) (Ramsay et al, 1988) show that, if primary tumors are treated upon reaching a diameter of 6 mm, the maximal percentage of metastasis is only 3.1% in FSall and 8.0% in SCCVII, while it increases to 14.3% and 41.3%, respectively, if the treatment starts at a diameter of 12 mm. When local recurrence of the same tumor lines are considered, the percentage of metastasis, at a tumor diameter of 6 mm, is only 12.5% in FSall and 43.0% in SCCVII, respectively, versus 46.6% and 70.3% at a diameter of 12 mm. Furthermore, the diameter doubling time (± standard deviation) necessary to grow from 6 to 12 mm increases from $\Delta t_p$=6.6 (± 0.4) days in primary to $\Delta t_r$=13.6 (± 4.6) days in recurrent Fsall tumors, and from $\Delta t_p$=7.7 (± 0.5) days to $\Delta t_r$=12.8 (± 3.8) days in SCCVII, respectively. An approximate computation from our model yields $\Delta t_r / \Delta t_p \sim 1.7$, which is in agreement with experimental results from Ramsay et al (1988). Hence, the recurrent cancer grows much slower than the primary tumor. Following the notion of a universal law, a possible explanation for this behavior is that the residual clonogenic cells of the primary tumor, which lead to recurrence, generate cells belonging to an 'older' developmental phase.

*b) cell turnover rates.* As another consequence of the model, the relative amount of energy devoted to tumor growth, $R = 1 - r$ can be related to the proportion $\Delta N/N$ of the cells contributing to the growth at any 'rescaled time', $\tau$:

$$R = \exp(-\tau) \approx \frac{\Delta N}{N} \qquad [4]$$

Here, $N$ is the total number of tumor cells and $\Delta N$ is the difference between the rates of newly generated tumor cells (growth fraction (GF)) and tumor cells being lost (i.e., cell loss factor (CLF), which includes both cell death and cell invasion). Eq. [4] predicts that $\Delta N$ is larger at the onset of tumor growth, and decreases exponentially afterwards. In fact, data from glioblastoma spheroids (Nirmala et al, 2001) suggest values for $\Delta N/N$ ranging from 0.9 at 1 week to 0.35 after 4 weeks. Moreover, knowing the rate of cell turnover at any developmental stage can help to improve therapeutic strategies. As such, we may define the 'lethal' dose of a given therapeutic agent as the dose *D* for which the rate of tumor growth post treatment, GF(D) equals the rate of CLF. GF(D) depends then on both the developmental stage of the tumor (i.e., *t*) and on its therapeutical response. If we further assume a given survival fraction SF(D) (different for each tumor), when a dose *D* has been administered, then GF(D) should become equal to the GF of the untreated tumor multiplied by SF(D). There is, however, a threshold time (corresponding to the *"critical"* tumor volume at inception of metastasis), before which the dose *D* is *curative*, while later it can only be *locally effective*. A longitudinal study of the metastasis-free survival in prostate carcinoma (Fuks et al, 1991) in 679 patients has shown that after 5, 10 and 15 years the percent survival of local control (LC) cases was 90%, 82% and 77%, respectively, while for local failure (LF) cases it was 68%, 39% and 24%, respectively. One could therefore argue that the LC tumors were correctly treated (i.e., at doses larger than the lethal one) early on and that only in a small percentage of patients was the metastasis threshold already surpassed at the time of treatment. Conversely, LF doses were inadequate and therefore, as the tumor mass continues to grow, metastases were inevitable.

c) *angiogenesis*. It is noteworthy that West 's model applies to organisms growing in *unrestricted* dietary conditions. Correspondingly, only fully replenished tumors should follow the universal growth curve. And in fact, as Freyer and Sutherland (1986) pointed out, differences in growth rates and saturation sizes (of up to a factor of 500) were found comparing tumor spheroids cultured in media with different oxygen and glucose concentrations. As such, deviations from the universal curve are conceivable depending on the particular environmental conditions. Our formalism may yield a simple mathematical explanation for such growth behavior. From the definition of *r* follows:

$$dr/dM = -\frac{1}{4} \cdot (m/M^5)^{0.25} < 0 \qquad [5]$$

likewise, from the expression of *M* as a function of the metabolic rate in West et al (2001) of a single cell $B_c$ we deduce that also

$$dM/dB_c < 0 \qquad [6]$$

Thus, if the average value of $B_c$ decreases, such as due to the depletion of a e.g. non-replenished nutrient concentration, the *r* curve is bound to shift downwards. Therefore, as "ontogenetic" tumor growth should depend on the availability of nutrient replenishment or, *in vivo*, on the successful competititon for available nutrients, any marked deviation between the universal and the *real* growth curve may indicate the amount of additional nutrients necessary to satisfy the metabolic demand of the growing tumor. Accordingly, one can further hypothesize that *in vivo* the degree of deviation from the universal curve may in fact relate to the amount of paracrine endothelial cell growth factors (e.g., VEGF (Plate et al, 1992; Shweiki et al, 1992) released by the nutrient-deprived tumor cells in order to induce the extent of *neovascularization* necessary to provide the required level of nutrient replenishment. It is noteworthy that the importance of nutrient availability has been pointed out also by a model proposed by Delsanto et al. (2000) and Scalerandi et al (1999, 2001) where, following the Local Interaction Simulation Approach (LISA) (Schechter et al, 1994), a consistent set of rules governing the microscopic interactions (at the cellular level) has been formulated, amenable to direct numerical simulations of cancer growth both in the avascular and vascular case.

d) *Invasion*. Finally, our findings also relate to *cell motility*, another hallmark of malignant tumors. Extracellular matrix-degrading invasive cells are likely to decrease the mechanical confinement around the tumor. These cells also appear to follow specific routes of dissemination, which include the perivascular space in the case of malignant astrocytomas (Bernstein et al, 1989; Vajkoczy et al, 1999). As such, at least in this case, tissue invasion should facilitate further growth by reducing the constraint of confinement not only for the volumetrically expanding tumor itself but also for the aforementioned chemo-attracted endothelial cells, which migrate towards the angiogenic factor releasing tumor cells. Thus, a progressive decrease in $B_c$ may not only trigger an increase in (e.g.) VEGF secretion, but could also induce an *increase* in the extent and dynamics of tissue invasion, which in turn facilitates angiogenesis and ensures continuing tumor growth. Therefore, not only metastasis and angiogenesis but also tumor cell invasion may be linked to a *"critical"* tumor volume, as has already previously been argued on the basis of experimental findings from human glioma spheroids (Deisboeck et al, 2001).

In summary, we have investigated the applicability of West's general model for the ontogenetic growth of living organisms to the case of solid malignant tumors. The results from fitting a variety of *in vitro* data from multicellular tumor spheroids and of *in vivo* data from both experimental models as well as patients support the notion that neoplasms also follow a universal growth law. Our paradigm-shifting finding has far-reaching implications concerning the mechanism of tumor metastasis and recurrence, cell turnover, angiogenesis and invasion.

.

**ACKNOWLEDGEMENTS:** The work has been partially supported by the 'Compagnia di S. Paolo', Torino, Italy. The authors would like to thank Dr. Yuri Mansury (Complex Biosystems Modeling Laboratory, Harvard-MIT (HST) Athinoula A. Martinos Center for Biomedical Imaging, Massachusetts Institute of Technology) for helpful suggestions.

**FIGURE CAPTIONS:**

**Fig. 1.** Tumor growth curves (replenished multicellular tumor spheroids). $r = (m/M)^{0.25}$ vs the 'rescaled' time $t$ for various multicellular spheroids. Rat gliosarcoma (9L) and human glioblastoma (U118) were cultured over 60 days on agarose with the medium superlayer routinely replaced every seven days (Chignola et al, 2000). The fitting lines were obtained by the same authors using a Gompertzian model. Data from another human glioblastoma line (SNB19) (Nirmala et al, 2001), cultured over a shorter time (30 days, with the cell culture medium replenished once every other day) and reaching a maximum volume of about 0.15 mm$^3$ are also included. In spite of the different dimensions (after 30 days, the 9L and U118 spheroids measured 0.5 mm$^3$ and 0.9 mm$^3$, respectively), all data fit well eq.[2], which is also plotted for reference.

**Fig. 2.** Tumor growth curves (rodent models). Fibroadenoma (Fibro) and Walker carcinoma (Walker) were transplanted into rats: their initial tumor volumes ranged from 0.1 (Walker) to 1 cm$^3$ (Fibro), and final volumes approached 100 cm$^3$ after 180 days and 25 days, respectively (Steel, 1977). The other murine tumors: a mammary adenocarcinoma (KHJJ), a mammary carcinoma (C3H), a sarcoma (EMT6) and two osteosarcomas (NCTC2472 and osteo) had initial volumes ranging between 0.001 and 0.05 cm$^3$ and final volumes of 1 to 5 cm$^3$, after an elapsed time between 15 and 100 days, depending on the tumor (Steel, 1977). Additional data on another murine mammary carcinoma (Ch3 ISS) (Cividalli et al, 2002) are also included. Eq [2] is again plotted for reference.

**Fig. 3.** Tumor growth curves (breast and prostate tumors from patients). Data are extrapolated using the Gompertzian curve parameters (Norton, 1988; Yorke et al, 1993)**.** For patient data the analysis is more difficult, since often only few observations of the same untreated tumor are available and the tumor's asymptomatic hystory is unknown. As in the previous figures, Eq [2] is plotted here for reference.

**TABLE 1: Estimates of the relevant parameters from Figures 1-3.** An asterix * in the second column denotes *in vivo* data. The values of the parameters $m_0$, $M$ and $a$ are estimated by means of the fitting procedure described after eq. [3]. $R^2$ represents the correlation coefficient between actual and fitted data.

| *Tumor* | *Reference* | $m_o(g)$ | $M(g)$ | $a\ (g^{0.25}/day)$ | $R^2$ |
|---|---|---|---|---|---|
| 9L (exp) | Chignola, 2000 | 0.000037 | 0.000515 | 0.102 | 0.69 |
| 9L (fitted) | Chignola, 2000 | 0.0000189 | 0.000515 | 0.114 | 0.99 |
| U118 (exp) | Chignola, 2000 | 0.000037 | 0.000812 | 0.084 | 0.64 |
| U118 (fitted) | Chignola, 2000 | 0.0000605 | 0.000823 | 0.084 | 0.98 |
| SNB19 | Nirmala, 2001 | 0.025 | 3 | 0.075 | 0.96 |
| fibro | Steel, 1977 * | 1 | 200 | 0.14 | 0.95 |
| walker | Steel, 1977 * | 0.348 | 150 | 1.54 | 0.97 |
| KHJJ | Steel, 1977 * | 0.0012 | 2 | 0.23 | 0.99 |
| C3H | Steel, 1977 * | 0.0348 | 5 | 0.21 | 0.98 |
| EMT6 | Steel, 1977 * | 0.00135 | 3 | 0.5 | 0.99 |
| NCTC 2472 | Steel, 1977 * | 0.052 | 7 | 0.49 | 0.99 |
| osteo | Steel, 1977 * | 0.0058 | 7 | 0.124 | 0.92 |
| C3H ISS | Cividalli, 2002* | 0.2 | 8 | 0.37 | 0.95 |
| human breast | Norton, 1988 * | 1 | 646 | 0.81 | 0.97 |
| human prostate | Yorke, 1993 * | 1 | 641 | 0.42 | 0.85 |

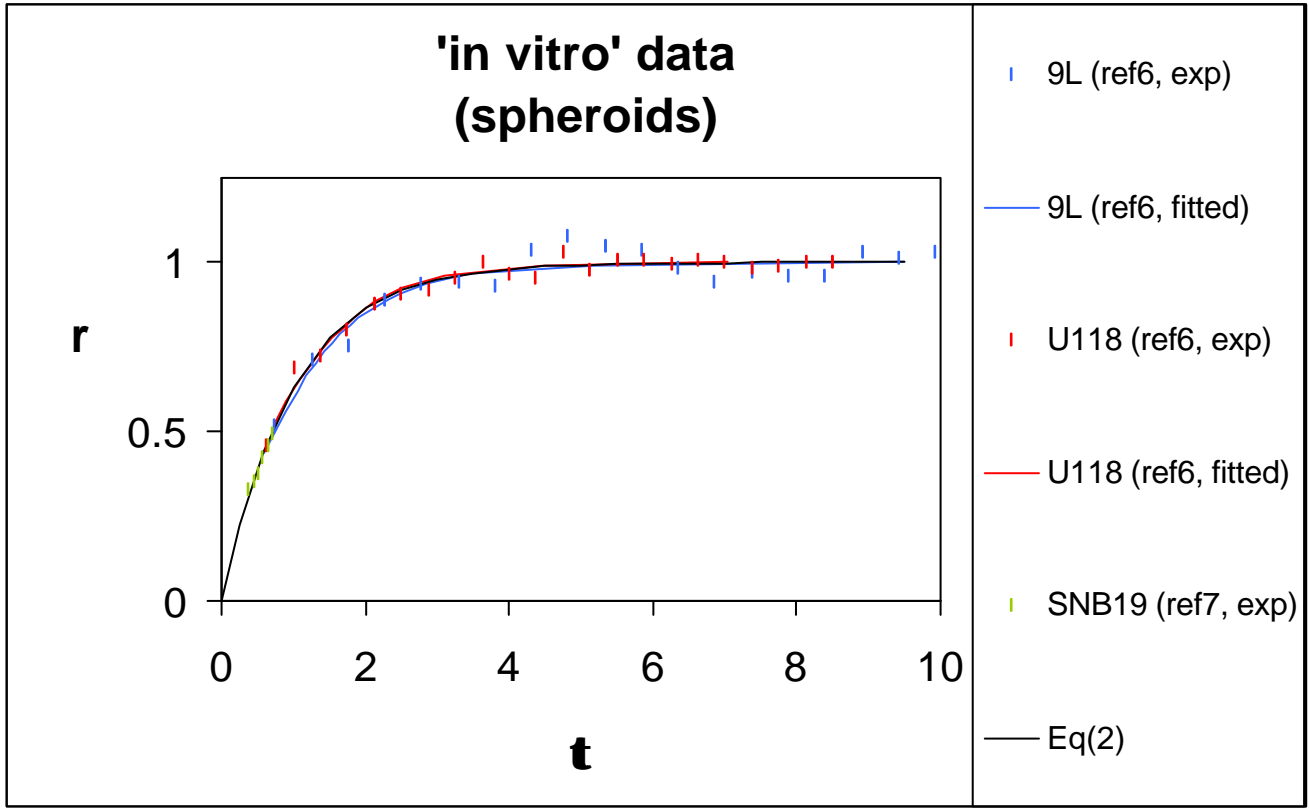

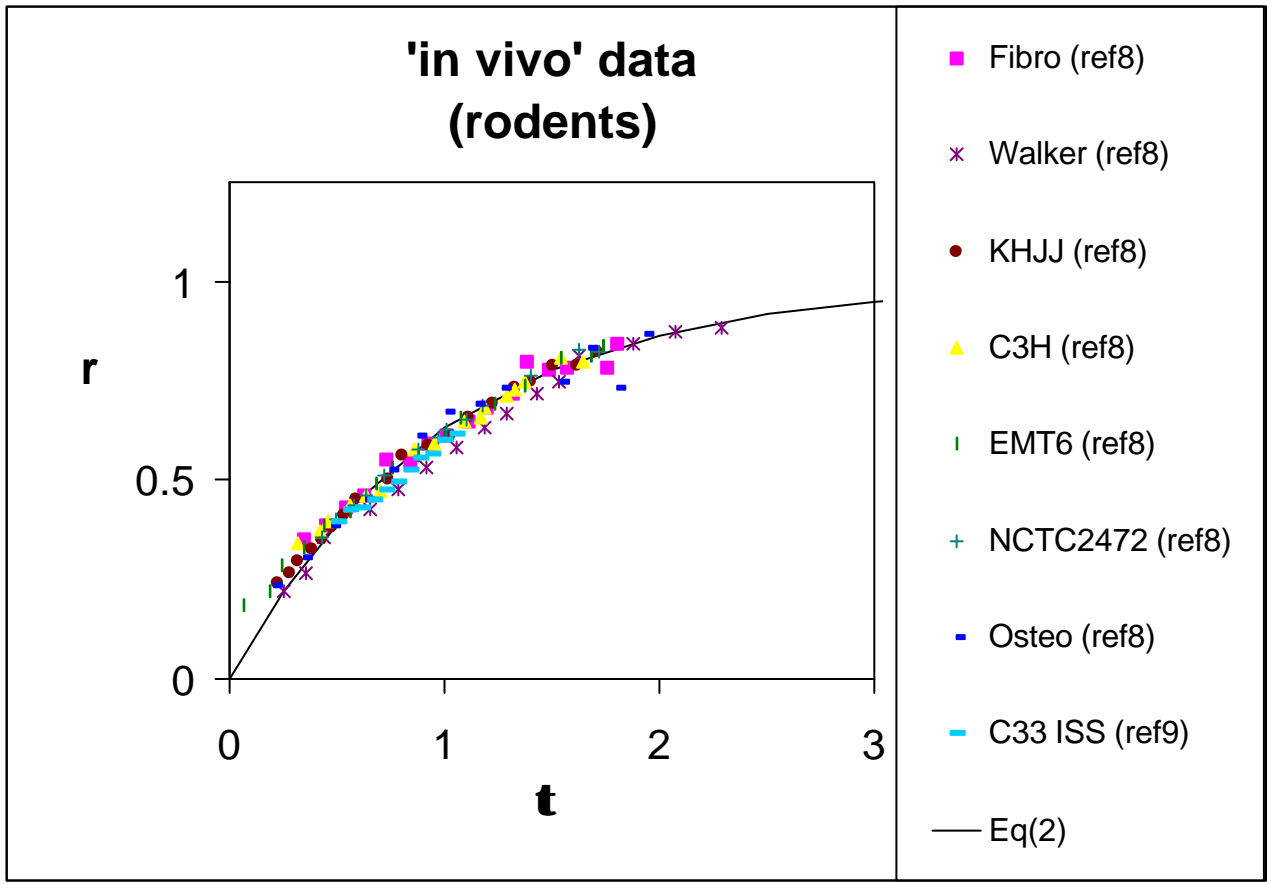

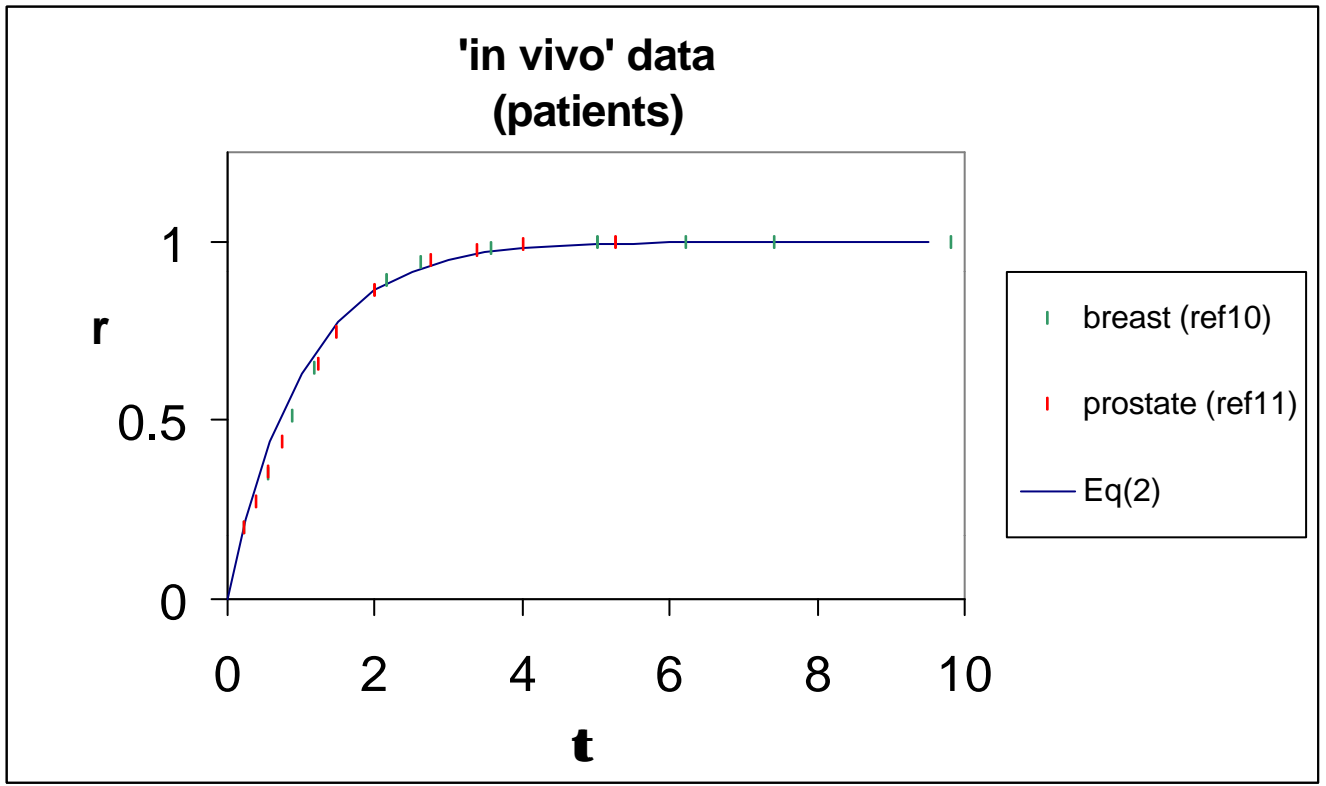